\documentclass{llncs}
\usepackage{hyperref}
\usepackage[bottom]{footmisc}

\usepackage{url}

\begin{document}

\title{Several Types of Types\\ in Programming Languages}

\author{Simone Martini}
%
\institute{Universit\`{a} di Bologna, Dipartimento di Informatica--Scienza e Ingegneria, Italy;\\ and INRIA Sophia-Antipolis, France
}
\maketitle

%
%

\begin{abstract}
Types are an important part of any modern programming language, but we often forget that the concept of type we understand nowadays is not the same it was perceived in the sixties. Moreover, we conflate the concept of ``type'' in programming languages with the concept of the same name in mathematical logic, an identification that is only the result of the convergence of two different paths, which started  apart with different aims. The paper will present several remarks (some historical, some of more conceptual character) on the subject, as a basis for a further investigation. We will argue that there are three different characters at play in programming languages, all of them now called types: the technical concept used in language design to guide implementation; the general abstraction mechanism used as a modelling tool; the classifying tool inherited from mathematical logic. We will suggest three possible dates \emph{ad quem} for their presence in the programming language literature, suggesting that the emergence of the concept of type in computer science is relatively independent from the logical tradition, until the Curry-Howard isomorphism will make an explicit bridge between them.
\\[1em]
\end{abstract}

\section{Introduction}
Types are an important part of modern programming languages, as one of the prominent abstraction mechanisms over data\footnote{Even in ``\emph{untyped}'' languages (Python, say) types are present and relevant.}. This is so obvious that we seldom realise that the concept of type we understand nowadays is not the same it was perceived in the sixties, and that it was largely absent (as such) in the programming languages of the fifties. Moreover, we now conflate the concept of ``type'' in programming languages with the concept of the same name in mathematical logic---an identification which may be (is it?) good for today, but which is the result of a (slow) convergence of two different paths, that started quite apart with different aims. Tracing and recounting this story in details, with the painstaking accuracy it merits, it is well beyond the limits of this paper---it could be the subject of a doctoral thesis. We will instead make several remarks (some historical, some of more conceptual character) that we hope will be useful as a basis for a further investigation. We will argue that there are three different characters at play in programming languages, all of them now called \emph{types}: the \emph{technical concept} used in language design to guide implementation; the \emph{general abstraction mechanism} used as a modelling tool; the \emph{classifying tool} inherited from mathematical logic. We will suggest three possible dates \emph{ad quem} for their presence in the programming language literature, suggesting that the emergence of the concept of type in computer science is relatively independent from the logical tradition, until the Curry-Howard isomorphism will make an explicit bridge between them.
As we will  see, the investigation on the arrival on the scene of these three characters will bring us to the (early) seventies.

\section{From types to ``types''}
One of the first questions to be cleared is when the very word ``\emph{type}'' stably entered the technical jargon of programming languages\footnote{Which is not to say when it was first used in that context. To our knowledge, the very first technical use of the term ``type'' in programming is H.B. Curry's~\cite{Curry1949}, to distinguish between memory words containing instructions (``\emph{orders}'') and those containing data (``\emph{quantities}''). These reports by Curry, as reconstructed by~\cite{DeMol2013}, contain a surprising and non-trivial mathematical theory of programs, up to a theorem of the style ``well-typed expressions do not go wrong''! Despite G.W. Patterson's review on JSL 22(01), 1957, 102-103, we do not know of any influence of this theory on subsequent developments of programming languages.
}. Contrary to folklore, early documentation on FORTRAN does not use the word, at least in the technical sense we mean today. In one of the early manuals, dating 1956~\cite{FORTAN56}, we read, for instance
\begin{quotation}\noindent
Two types of constants are permissible: fixed points (restricted to integers) and
floating points {\tiny (page 9)}
\end{quotation}
or
\begin{quotation}\noindent
Two types of variables are also permissible (with the usual distinction based
on the initial letter) {\tiny (page 10)}
\end{quotation}
but also
\begin{quotation}\noindent
32 types of statement {\tiny (page 8)}
\end{quotation}
These are generic uses of the term ``type''---``kind'' or ``class'' could be used instead. 
Especially because, on page 14 there is a detailed discussion of what happens
when one mixes integers and floats. And ``type'' is never used. The noun ``mode''
is  used instead\footnote{
Of course the distinction between integers and floating points---that is, a type-based distinction, in today's terminology---was present and used, to decide the memory layout of the different kinds of variables, and to compile into the correct arithmetic operations.
}:
\begin{quotation}\noindent
A FORTRAN expression may be either a fixed or a floating point expression, but it must not be a mixed expression. This does not mean that a floating point quantity can not appear in a fixed point expression, or vice versa, but rather that a quantity of one mode can appear in an expression of the other mode only in certain ways. (\ldots) Any fixed point (floating point) constant, variable, or subscripted variable is an expression of the same mode. (\ldots) If SOMEF is some function of n variables, and if E, F, \ldots ,H are a set of n expressions of the correct modes for SOMEF, then SOMEF (E, F,
\ldots , H) is an expression of the same mode as SOMEF. {\tiny (page 14)}
\end{quotation}

When, then, do we find a precise occurrence of our technical term? For sure in the report on Algol 58~\cite{Algol58}
published in December 1958. There, ``type'' is used as a collective representative for ``special types, e.g., \emph{integral}, or \emph{Boolean}'' (page 12). Declarations (needed for non real-valued identifiers) are called ``type declarations'':
\begin{quotation}\noindent
Type declarations serve to declare certain variables, or functions, to represent quantities of a
given class, such as the class of integers or class of Boolean values.\\
Form: $\Delta \sim$ \emph{type} (I,I,\ldots I) where \emph{type} is a symbolic
representative of some type declarator such as \emph{integer} or \emph{boolean}, and the I are identifiers. Throughout the program, the variables, or functions named by the identifiers I, are constrained to refer only to quantities of the type indicated by the declarator {\tiny (page 16)}.
\end{quotation}
Algol 58 is the result of a meeting held in Zurich at the end of May 1958, between an ACM group (including Backus and Perlis) and a European group.
Each group had its own preparatory paper~\cite{ACMadhocAlgol1958,Bauer1958}, and both such papers \emph{do not} use ``type''. Of the two, only the ACM's one discusses the issue of the declaration for non real-valued identifiers, using the general term ``class'': 
\begin{quotation}\noindent
A data symbol falls in one of the following classes:
a) Integer b) Boolean c) General {\tiny (page 4)}
\end{quotation}
Declarations are called ``Symbol Classification Statements''; their concrete syntax is the same 
as in the final Algol 58 report\footnote{Recall that \emph{type} is not a reserved word in  Algol 58---it is used in the report for the ``symbolic
representative of some type declarator such as'' INTEGER, BOOLEAN, etc.}:
\begin{quotation}\noindent
The symbol classification statements are:\\
INTEGER ($s_1,\ldots,s_n$) \\
BOOLEAN ($s_1,\ldots,s_n$)
\end{quotation}
but it is striking how during the Zurich meeting the committee realised that the different ``classes'' could be grouped together, and given a name as a collective---types were born. It is also remarkable that, at least from these references, the technical term appears to be just a semantical shift from the generic one; in particular, there is no clue that in this process the technical term ``type'' from mathematical logic had any role\footnote{Alan Perlis summarises in 1978, referring to Algol 58: ``The use of `type,' as in `x is of type {\bf real},' was analogous to that employed in logic. Both programming language design and logic dipped into the English language and came up with the same word for roughly the same purpose''~\cite{Perlis1981}.
}. This process will come to  maturity in Algol 60~\cite{Algol60}:
\begin{quotation}\noindent
Integers  are of type {\bf integer}. All other numbers are of type {\bf real}.
\end{quotation}
or
\begin{quotation}\noindent
The various ``types''
({\bf integer}, 
{\bf real}, 
{\bf Boolean}) 
basically denote properties of values. 
\end{quotation}
Observe the word ``types'' under quotes, as to stress that it is no longer the ordinary word, but the technical one.

What this term means is simple---data values are partitioned in disjoint classes; each class is mapped to a specific memory representation. Type information collected from the source program guides the compiler for memory allocation and choice of machine operations to be used during translation. Moreover, these types provide a certain level of abstraction over such implementation details, avoiding the manipulation of a value by operations of different types. However, besides the availability of indexed values (arrays), there is no linguistic provision for dealing with more structured data, or for data ``naturally'' belonging to classes not available as primitive types. 

\section{Data types and abstractions}
Algol 58 treats arrays separately from types. One first declares the type of an identifier (unless it is a real-valued one, for which no such declaration is needed), than declares the identifier to be an array, fixing the number of dimensions (and assigning lower and upper bounds for the indexes). With all its maturity with respect to ``types'', Algol 60 makes no change in the treatment of arrays---types denote properties of just ``simple'' values. 

That Algol's provision for primitive data was too restrictive, was clear even to its designers\footnote{E.g., ``ALGOL (\ldots) lacks the ability to describe different kind of data''~\cite{McCarthy1961} (note that once again the generic ``kind'' is used, and not ``type''). Cfr also~\cite{Priestley2011}, page 244.}. To address this ``weakness,''
John McCarthy advocates a
\begin{quotation}\noindent
way of defining new data spaces in terms of given base spaces and of defining functions on the new spaces in terms of functions on the base spaces.~\cite{McCarthy1961}. {\tiny (page 226)}
\end{quotation}
The new data space constructors are the Cartesian product, the disjoint union, and the power set, each of them equipped with its canonical (universal) maps, which are used to define functions on the new spaces from functions on the base spaces. McCarthy's paper treats the question at a general meta-level, it does not propose a specific language, and it does not use the term ``type'', but it sets a clear roadmap on how to introduce new types in programming languages---instead of inventing an arbitrary new palette of primitive types, provide general, abstract\footnote{Category theory and Bourbaki are clearly at an arm's length, but there is no explicit reference to them in the paper.} 
mechanisms for the construction of new types from the base ones. Base types could be taken as frugal as the single ``null set'', since natural numbers could be defined from it. Although McCarthy's paper has no explicit reference to any type-theoretic, mathematical logic paper (it cites Church's logic manual, though), we think this is one of the first contacts of the two concepts we are concerned with in this essay, albeit still in an anonymous form.

The challenge to amend the ``weakness of Algol'' was taken up in more concrete forms, and in similar ways, by Tony Hoare~\cite{Hoare1965}, and by Ole-Johan Dahl and Kristen Nygaard~\cite{Dahl:1966}, around 1965. Hoare's paper, with an explicit reference to McCarthy's project introduces at the same time the concepts of (dynamically allocated) \emph{record} and \emph{typed reference}.
A record
is an ordered collection of named \emph{fields}\footnote{It is a ``structure,'' in C's terminology.}; the life of a record does not follow the life of the block in which the record is created. 
Typed references may be seen like pointers, but no operations are allowed on them, besides creation and dereferencing (that is, access to the ``pointed'', or referenced, object). Moreover, when such a reference is created, the type (or class, in the paper's terminology) of the referenced record is fixed and cannot be dynamically modified. 
Records are not a new idea---the concept was introduced in ``business oriented languages'', FLOWMATIC first, then COBOL (see, e.g.,~\cite{Hopper1959}), where the field of a record may be a record itself (nested records), thus permitting static hierarchical structures (i.e., trees). 
Also dynamically allocated  structures\footnote{More precisely: dynamically allocated structure which do not follow a stack-based life policy.}
were already available in ``list processing languages'', of which LISP is the main representative. Lisp's \emph{S-expressions}~\cite{McCarthy:1960} may indeed be seen as dynamic records composed of two unnamed fields. Moreover, since S-expressions may be nested, they may be used to simulate more complex structures. 
What is new in Hoare's proposal, however, is from one side the flexibility in modelling provided by arbitrary named fields; from the other, and crucially, Hoare's records may contain references to other records, thus allowing for the explicit representation of graph-like structures. 

In Simula~\cite{Dahl:1966}, Dahl and Nygaard had already implemented analogous ideas, with the aim to design an extension to Algol for discrete event simulation: a record class is an \emph{activity}; a record is a \emph{process}; a field of a record is a local variable of a process (see also~\cite{Hoare1966}). References are not a prime construct of the language; instead, there are \emph{sets}, which are bidirectional lists of \emph{elements}, each of them being (a pointer to) a process. What is really new in Simula I is that a (dynamically created) ``process'' encapsulates both data objects and their associated operators, a concept that will be called \emph{object} in Simula 67 (see, e.g.,~\cite{Dahl2001}) and which will be popularised by Alan Kay in the context of Smalltalk~\cite{Smalltalk1976,Kay:1993}. 

Of the two papers we are discussing, it will be Hoare's one to have the major, immediate impact. Although the proposal is for an extension to Algol 60, it will not materialise into the ``official'' Algol family---Algol W, which we shall discuss later,
is not an official product of the Algol committee\footnote{Hoare's paper will have significant impact also on Algol 68---the legitimate child of the Algol committee---which contains references and structured types. 
Tracing the genealogy of Algol 68's \emph{modes} (Algol 68's terminology for types) is however a task that should be left for the future.
}.
The paper is fundamental because types change their ontology---from an implementation issue, they programmatically become a general abstraction mechanism\footnote{
In  John Reynolds's words from 1983, ``Type structure is a syntactic discipline for enforcing levels of abstraction''~\cite{Reynolds1983}. Or in those of Luca Cardelli and Peter Wegner from their seminal 1985 paper, ``The objective of a language for talking about types is to allow the programmer to name those types that correspond to interesting kinds of behavior''~\cite{CardelliWegner85}.
}:
\begin{quotation}\noindent
the proposal is no arbitrary extension to an existing language, but represents a genuine abstraction
of some feature which is fundamental to the art or science of computation. {\tiny (page 39)}
\end{quotation}
This happens on (at least) three levels. First, it implements McCarthy's project into a specific programming language, extending the concept of type from simple to structured values\footnote{From the terminological point of view, the paper uses ``classes'' when referring to records, and ``types'' for simple types (integer, real, boolean \emph{and} references, which are typed: the type of a reference includes the name of the record class to which it refers). On page 48, however,  discussing the relations with McCarthy's proposal, we find cristal-clear awareness: ``The current proposal represents part of the cartesian suggestion made by Prof. J. McCarthy as a means of introducing new types of quantity into a language.'' 
From Hoare's expression ``record class'', Dahl and Nygaard derive the term ``object class'' in Simula 67~\cite{Dahl2001}, then simply ``class'' in the object oriented jargon.
}. Starting from this paper,  ``structured values'' are organised in types in the same way as ``simple values'',
thus opening the way to the modern view of \emph{data types}. 

Second, types are a linguistic modelling tool:
\begin{quotation}\noindent
In the simulation of complex situations in the real world, it is necessary to construct in the computer analogues of the objects of the real world, so that procedures representing types of even may operate upon them in a realistic fashion. {\tiny (page 46)}
\end{quotation}
The availability of a flexible way of data structuring (contrasted to the rigid structure provided by arrays) is seen as the linguistic mechanism that provides the classification of ``the objects of the real world''. Moreover, the possibility to embed references into records allows for the construction of complex relational structures. Data are no longer ``coded'' into integers or reals---a record type naturally represents a class of complex and articulated values. Even more importantly, following McCarthy, the language only provides general means of construction---the definition of  new classes of data being left to the programmer. 

Finally, the combination of record types and typed references provides a robust abstraction over the memory layout used to represent them. By insisting that references be typed, the type checker may statically verify that the field of a record obtained by dereferencing is of the correct type required by the context---primitive types are true abstractions over their representation. 
In retrospect,
\begin{quotation}\noindent
I realised that [types] were essential not only for determining memory requirements, but also for avoiding machine-dependent error in a running object program.  It was a firm principle of our implementation that the results of any program, even erroneous, should be comprehensible without knowing anything about the machine or its storage layout.~\cite{Hoare2014}
\end{quotation}
Hoare's proposal, including the terminology (``record classes''), will find its context into the joint paper~\cite{HoareWirth:1966}, and finally will be implemented in Algol W~\cite{AlgolW}, which will have a significant impact on subsequent languages, being an important precursor of Pascal. 
In Algol W the picture and the terminology are complete:
\begin{quotation}\noindent
Every value is said to be of a certain type. 
(\ldots) 
The following types of structured values are distiguished: array: (\ldots), record: (\ldots). {\tiny (pages 16-17)}
\end{quotation}

The last step yet to be done was the unification of two aspects that were still distinct in Hoare's proposal---classification (i.e., modelling) and abstraction. In Algol W, primitive types and user defined record types do not enjoy the same level of abstraction.
On one hand, primitive types (integers or floats, say) are an opaque capsule over their implementation-dependent representation, and the type system ensures that on a primitive type only operations of that type are allowed. On the other hand, the user may well define a record class for modelling `the objects of the real world'', but there is no way of fixing which operations are allowed on such class, besides the general ones manipulating records and references. The user will probably define functions taking as argument values of these record classes, but the type system cannot enforce that \emph{only} such operations are allowed to manipulate those values. 
In the literature of the early seventies there are several proposals for allowing (and enforcing) stricter checks.
Morris~\cite{Morris:1973} advocates that the type system (including user-defined types) guarantee that only the prescribed operations on a type could operate on its values (forbidding thus the manipulation of the representations of those values). 
A thesis which will be further elaborated and formulated in modern terminology\footnote{Morris talks about ``protection,'' ``authentication'', ``secrecy''.} by Reynolds in his seminal~\cite{Reynolds:1974}, which also extends it to polymorphic situations:
\begin{quotation}\noindent
The meaning of a syntactically-valid program in a ``type-correct'' language should never depend upon the particular representation used to implement its primitive types. (\ldots) The main thesis of [Morris \cite{Morris:1973}] is that this property of representation independence should hold for user-defined types as well as primitive types.~
\end{quotation}
From now on, types will be the central feature of programming languages as we understand them 
today\footnote{The story of abstract data types, their relation to polymorphism, and how their parabola gives way to object oriented programming, is something to be told in a different paper, see~\cite{MartiniCiE2016}.
}.
\section{Classifying values}
Types inhabit mathematical logic since the early days, with the role of restricting the formation of  formulas, in order to avoid paradoxes\footnote{This is not the place where to discuss the emergence and the evolution of the concept of type in logic---we will limit ourselves to a single glimpse on the view of Russell and Whitehead, which will be the dominant one in the twentieth century. Stratification, or classification, in types, orders, or similar ways was already present in the nineteenth century, see, for instance, Frege's \emph{Stufe} (in the \emph{Grundgesetze}; before he also used \emph{Ordnung}), usually translated with ``level'', or ``degree''.
}. They are a discipline for (statically---as we would say today) separating formulas ``denoting'' something from formulas that ``do not denote''. In the words of the Preface to \emph{Principia Mathematica}~\cite{RussellWhiteheadPM}:
\begin{quotation}\noindent
It should be observed that the whole effect of the doctrine of types is negative: it forbids certain inferences which would otherwise be valid, but does not permit any which would otherwise be invalid.
\end{quotation}
The opposition ``denoting'' vs.\ ``non denoting'' becomes, in programming languages, ``non producing errors'' vs.\ ``producing errors''\footnote{``Well-typed expressions do not go wrong.''~\cite{Milner1978}
}.
Types as a classifying discipline for programs---and with the same emphasis on the fact that some valid formulas will be necessarily forbidden, for decidability's sake---are found in the programming languages literature as early as in the PhD thesis of Morris~\cite{Morris1968}:

\begin{quotation}\noindent
We shall now introduce a type system which, in effect,
singles out a decidable subset of those wfes that are safe; i.e., cannot given rise to ERRORs. This will disqualify certain wfes which do not, in fact, cause ERRORS and thus reduce the expressive power of the language.~{\tiny(page 89)}
\end{quotation}

Morris performs his ``analysis'' by taking first the type-free $\lambda$-calculus, and imposing then the constraints of the ``simple'' functional types, formulated as a type-assignment system. More specifically, Morris says that ``the type system is inspired by Curry's theory of functionality'', quoting~\cite{CurryFeys1958}, while there is no reference to~\cite{Church1940}, which apparently would have been a more precise reference. The reason could be that Church formulates his theory directly with typed terms, instead of seeing types as predicates on type-free terms. Were this the reason, Morris' thesis would be the first reference to the now common distinction between typing ``\`a la Curry'' and  ``\`a la Church''. 

Are these the types of mathematical logic? They share the same aims, but the connection is implicit, even unacknowledged. The fact that Church's~\cite{Church1940} is not cited by Morris could certainly be explained as we argued above, but it is nonetheless revealing of the lack of awareness for the mathematical logic development of the concept. The first explicit connection we know of, in a non technical, yet explicit, way is~\cite{Hoare1972}, but the lack of acknowledgement is going to persist---neither Morris'~\cite{Morris:1973} or Reynolds'~\cite{Reynolds:1974} cites any work using types in logic. Certainly the \emph{Zeitgeist} was ripe for the convergence of the two concepts, and there was a formidable middleman---$\lambda$-calculus. Used first by Landin as a tool for the analysis of Algol (and then by Scott, Strachey, Morris, Reynolds, and all the rest), at the dawn of the seventies $\lambda$-calculus was the \emph{lingua franca} of conscious programming language theorists, both in the type-free and the typed version. Programming languages and proof-theory were talking the same language, but the conflation was always anonymous. In Reynolds's \cite{Reynolds:1974} a second order (``polymorphic'') typed lambda-calculus is independently introduced and studied, almost at the same time in which Girard \cite{Girard1971} uses it as a tool to prove cut-elimination for second order logic;  
Milner \cite{Milner1978} presents a type-reconstruction algorithm for simple types, independently from Hindley~\cite{Hindley1969} (which will be cited in the final version). The Curry-Howard isomorphism~\cite{Howard1980} (the original manuscript dates 1969 and was widely circulated, at least in the proof-theory and lambda-calculus communities) will be the catalyst for the actual recognition\footnote{For a lucid account of the interplay between types, constructive mathematics, and lambda-calculus in the seventies, see~\cite{Cardone2009}, Section 8.1.
}, which comes only in Martin-L\"of's~\cite{MartinLof1982}, written and circulated in 1979, which presents a complete, explicit correspondence between proof-theory and functional languages. The paper will have significant impact on following research (and not only the one on programming languages).

This slow mutual recognition of the two fields tells a lot on their essential differences. 
For most of the ``types-as-a-foundation-of-mathematics'' authors,
types where never supposed to be actually used by the working mathematician
(with the debatable exception of Russell himself). It was sufficient that
\emph{in principle} most of the mathematics could be done in typed languages, so that paradoxes could be avoided.

Types in programming languages, on the contrary, while being restrictive in the
same sense, are used everyday by the working computer programmer. 
And hence, from the very beginning in Algol, computer science had
to face the problem to make types more ``expressive'', and ``flexible''\footnote{
See, for instance, the Introduction to~\cite{Milner1978} which calls for polymorphism to ensure flexibility. Damas-Milner~\cite{LCF1979} type inference provides a powerful
mechanism for enforcing type restrictions while allowing more liberal (but principled)
reasoning.}. If in proof-theory ``typed'' means first of all ``normalizing'', in computer science there are --- since the beginning --- well-typed programs which diverge. While mathematical logic types are perceived as constraints (they ``forbid'' something, as  in Russell's quote above), types in programming languages are experienced as an enabling feature, allowing simpler writing of programs, and, especially, better verification of their correctness\footnote{
This emphasis on the moral need for a programming language to assist (or even guide) the programmer in avoiding bugs or, worse, unintended behaviour in a program, is the core of what Mark Priestly~\cite{Priestley2011} identifies as the ``Algol research program'', a way of thinking to the design of programming languages which still today informs   most work in programming language research.
}. 

The crucial point, here and in most computer science applications of
mathematical logic concepts and techniques, is that computer science never used 
ideological glasses (types per se; constructive mathematics per se; linear logic per se; etc.),
but exploited what it found useful for the design of more elegant,
economical, usable artefacts.  This eclecticism (or even anarchism, in the sense of epistemological theory) is one of the distinctive traits of the discipline, and one of the reasons of its success.

But this is the subject of an entirely different paper.

\section*{Acknowledgments}
I am happy to thank Gianfranco Prini for helpful discussions (and for his---alas, remote in time---teaching on the subject).

\small
\bibliography{IFES}

\end{document}